\documentclass[a4paper, 11pt, notitlepage, twocolumn, superscriptaddress]{revtex4-1}
\usepackage{amsmath}
\usepackage{times}
\usepackage{graphicx}
\usepackage{setspace}
\usepackage{color}
\setcitestyle{authoryear,round,comma} 

\begin{document}
\singlespacing

\title{Calling Dunbar's Numbers}
\author{P. MacCarron}
\affiliation{SENRG, Department of Experimental Psychology, University of Oxford, OX1 3UD, England}
\author{K. Kaski}
\affiliation{Department of Computer Science, Aalto University School of Science, P.O. Box 15500, Espoo, Finland}
\author{R.I.M. Dunbar}
\affiliation{SENRG, Department of Experimental Psychology, University of Oxford, OX1 3UD, England}
\affiliation{Department of Computer Science, Aalto University School of Science, P.O. Box 15500, Espoo, Finland}

\begin{abstract}
The social brain hypothesis predicts that humans have an average of about 150 relationships at any given time. Within this 150, there are layers of friends of an ego, where the number of friends in a layer increases as the emotional closeness decreases.
Here we analyse a mobile phone dataset, firstly, to ascertain whether layers of friends can be identified based on 
call frequency. 
We then apply different clustering algorithms to break the call frequency of egos into clusters and compare the number of alters in each cluster with the layer size predicted by the social brain hypothesis. 
In this dataset we find strong evidence for the existence of a layered structure.
The clustering yields results that match well with previous studies for the innermost and outermost layers, but for layers in between we observe large variability.
\end{abstract}

\maketitle

\section{Introduction}

In recent years the availability of communication data has allowed us to analyse the nature of human relationships and interactions on a much larger scale than previously available (see, for example, \cite{onnela2007structure}). Although modes of communication have changed however, our brain sizes have not, and it is suggested there is a cognitive constraint on the number of face-to-face social interactions one may have \citep{dunbar1993coevolution,roberts2009exploring}. 
This constraint fits in a broad sense with the 
`social brain hypothesis' which argues that the evolution of primate brains was driven by the need to maintain increasingly large social groups \citep{humphrey1976social,Dunbar1992,barton1979,Dunbar1998}. 

Individuals do not give equal weight to each relationship and evidence from the social brain hypothesis suggests that ego networks are structured into a sequence of layers with the size of each layer increasing as emotional closeness decreases \citep{Dunbar1998,hill2003social}. 
The mean number of friends in each 
has been found to be around 5, 15, 50 and 150 in the cumulative layers (i.e. on average 10 people in the second layer to make a total of 15) \citep{zhou2005discrete,hamilton2007complex}. Beyond this there are even larger groupings suggested at 500 and 1,500 \citep{dunbar1993coevolution,zhou2005discrete}. 

Recently these Dunbar layers have been observed in online social media, such as Facebook and Twitter \citep{dunbar2015structure} and an online computer game \citep{fuchs2014fractal}. 
These relationships are temporal 
 and so the 150 in particular
represents the amount of friends at a given time.
If a new friend is made, an old one is most likely dropped, and the strength relationships changes 
quicker in the outer layers than the inner ones \citep{sutcliffe2012relationships,saramaki2014persistence}.
However, other methods for estimating personal network sizes have found numbers larger than the outer Dunbar layer, these studies suggest an average personal network size of around 290 for Americans \citep{killworth1984,mccarty2001}.

Here we use a mobile phone call dataset initially 
to ascertain whether layers of friends are detectable in an offline context. If we find evidence of these layers, we then test if they match the layer sizes previously identified using different clustering algorithms.

A European phone-call dataset 
over all 12 months of 2007 is used. This has
34.9 million
users with almost 6 billion calls.
About 6 million of these users are with the company (who provide coverage to approximately 20\% of the country's population) for whom we have data on all calls they make.

The call frequency between two individuals represents the strength of a relationship and has been shown to correlate with emotional closeness \citep{roberts2011costs,arnaboldi2013egocentric}.
\cite{saramaki2014persistence} have also shown that social signatures in cellphone data remain robust over time even with identity changes in the alters.

\section{Methods}

To eliminate casual calls and business calls, the data are filtered so that only calls are counted if there is at least one reciprocal call between the two users. 

People vary in the extent to which they use their phones, with some using it as a regular means of communication with family and friends, and others using it only for social emergencies or to arrange meetings. While the former are likely to provide a full coverage of their social network, the latter won't. To avoid this kind of under-reporting, we censored the dataset so as to include only those individuals with a minimum number of alters. Since the average number of alters at a given time in personal, or ego-centric, networks is 150, with a natural range of 
approximately 100-250 \citep{hill2003social,zhou2005discrete,roberts2009exploring}, 
we initially set a value of 100 alters as the minimum cut-off. By doing so, we aimed to have a more complete distribution of actual ego networks, while not biasing against individuals who have naturally small networks. 
After this we lower the cut-off to 50 alters to observe the results for lower frequency users.

The degree $k$ of an ego represents the number of alters called and the weighted degree $w$ represents the total number of calls an ego makes. 
The degree distribution $p_k$ and weighted distribution $p_{w}$ are the fraction of vertices in a network with degree $k$ and weighted degree $w$, respectively. Note that in 
empirical networks, the degree distributions are often found to have positive or right skew \citep{Newman2003}.  

In order to estimate the functional forms of degree distribution, the method of Maximum Likelihood Estimators is used \citep{Clauset,Edwards}. Here we test different distributions; namely power law, exponential, stretched exponential, Gaussian (or normal) and log-normal distributions, and use the Akaike Information Criteria to select the best model \citep{Akaike,Burnham}. 

The data for each user is considered as a one dimensional array which we denote by $W$ such that 
the minimum possible weight is $w_{\rm{min}} = 1$ when an ego calls an alter once. There is no real upper limit (beyond financial or time constraints) to the maximum number of calls a user can make to their preferred alter. 
In order to compare users, the data for user $i$ is normalised by 
\begin{equation}
\hat{W_i} = \frac{w_i - w_{i \text{ min}}}{ w_{i\text{ max}}- w_{i\text{ min}}},
\label{eqn:norm}
\end{equation} 
where $w_i$ is the number of calls made to each alter and $w_{i\text{ min}}$ and $w_{i\text{ max}}$ are minimum and maximum number of calls they make to any of their alters. 
This ensures that, for each ego, the strongest interaction with an alter is 1 and the weakest is 0.
A first estimate to identify the layers is to plot the probability density of all different weights for all users to ascertain if any pattern exists.  A kernel density estimate is applied to the true probability density and the local minima are used to identify clusters \citep{rosenblatt1956remarks,parzen1962estimation}. 

Many methods exist for data clustering, (see, for example, \citealt{jain1999data}, \citealt{gan2007data}). The vast majority of these algorithms, however, are for high-dimensional datasets \citep{jain2010data}. Here, although we are dealing with big data, we seek to break each individual's calls into clusters or layers. Thus we are dealing with one-dimensional clustering for each user, and from this we analyse the average layer sizes.

A common method for one-dimensional clustering is the Jenks natural breaks algorithm \citep{jenks1967data}. 
The Jenks algorithm is similar to $k$-means clustering in one dimension \citep{khan2012initial}. It searches for the minimum distance between data points and the centres of the clusters they belong to as well as for maximum difference between cluster centres themselves.
The goodness of fit can be calculated to optimise the number of clusters found 
\citep{coulson1987matter}.
A goodness of fit of 1.0 can only be attained when there is zero within-class variation (often when the number of clusters is the same size as the data). To choose the optimal number of clusters 
we take a threshold of 0.85 for the goodness of fit as suggested in \cite{coulson1987matter}.

We also use a Gaussian Mixture Model which assumes that the data are generated from a number of Gaussian distributions \citep{day1969estimating}. Naively, we may assume that the layers are made up of Gaussian distributions with their means on the Dunbar numbers. The expectation maximisation algorithm is implemented for this \citep{dempster1977maximum} and, again, the Akaike Information Criterion is used to assess the number of clusters in the data.

Another method for clustering the data, used here, is the head/tail breaks \citep{jiang2013head}. This method was developed for data with heavy-tailed distributions. It splits the data at the mean and taking the head (all values above the mean), it recursively splits each consecutive head at its mean. Our data is heavy tailed \citep{onnela2007structure}, with most users calling many people a small number of times but calling their closer friends frequently. An advantage of the head/tail breaks is that the number of clusters is derived naturally from the distribution of the data. 

\begin{figure*}[t!]
\begin{center}
\includegraphics[width=0.99\columnwidth]{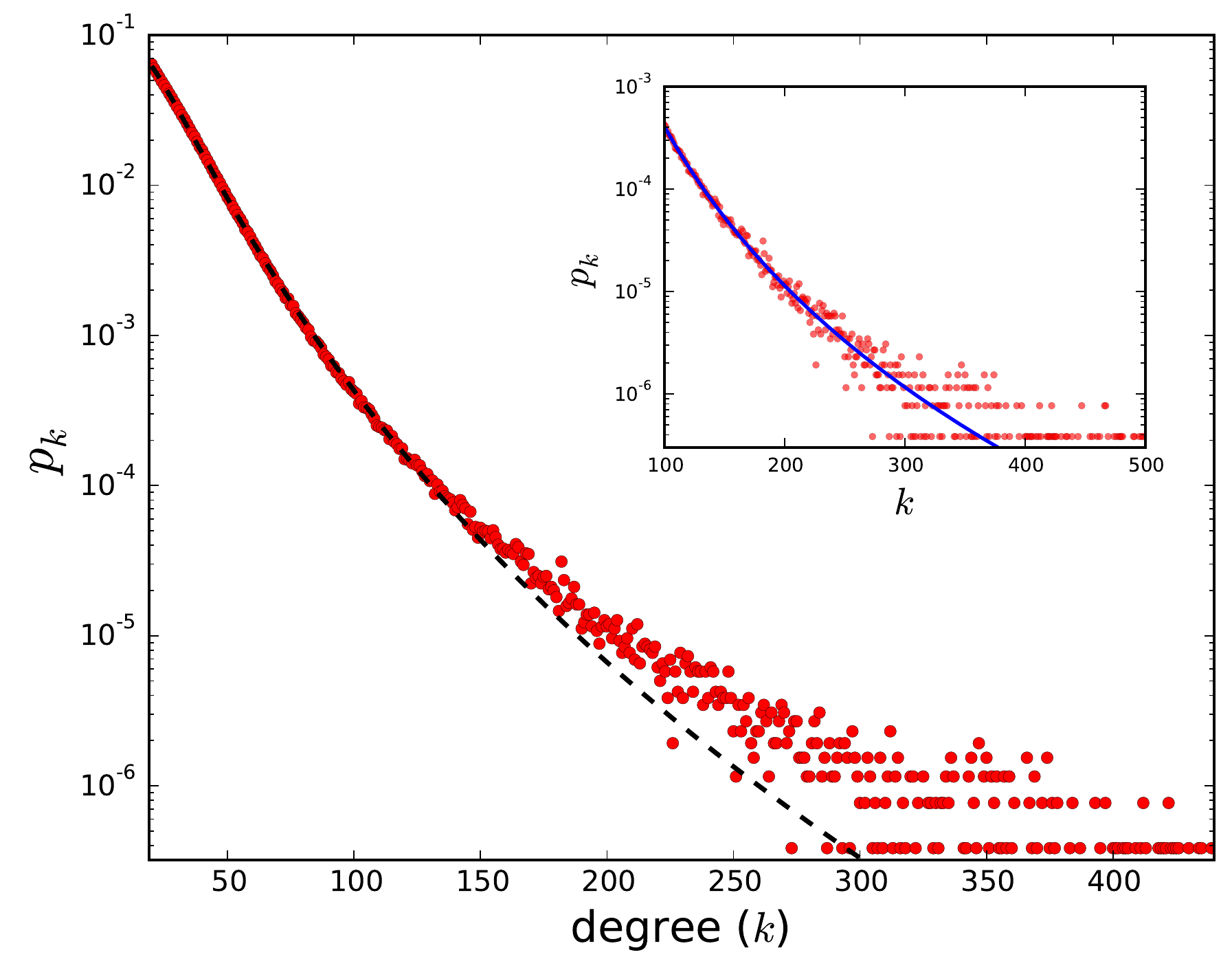}
\includegraphics[width=0.99\columnwidth]{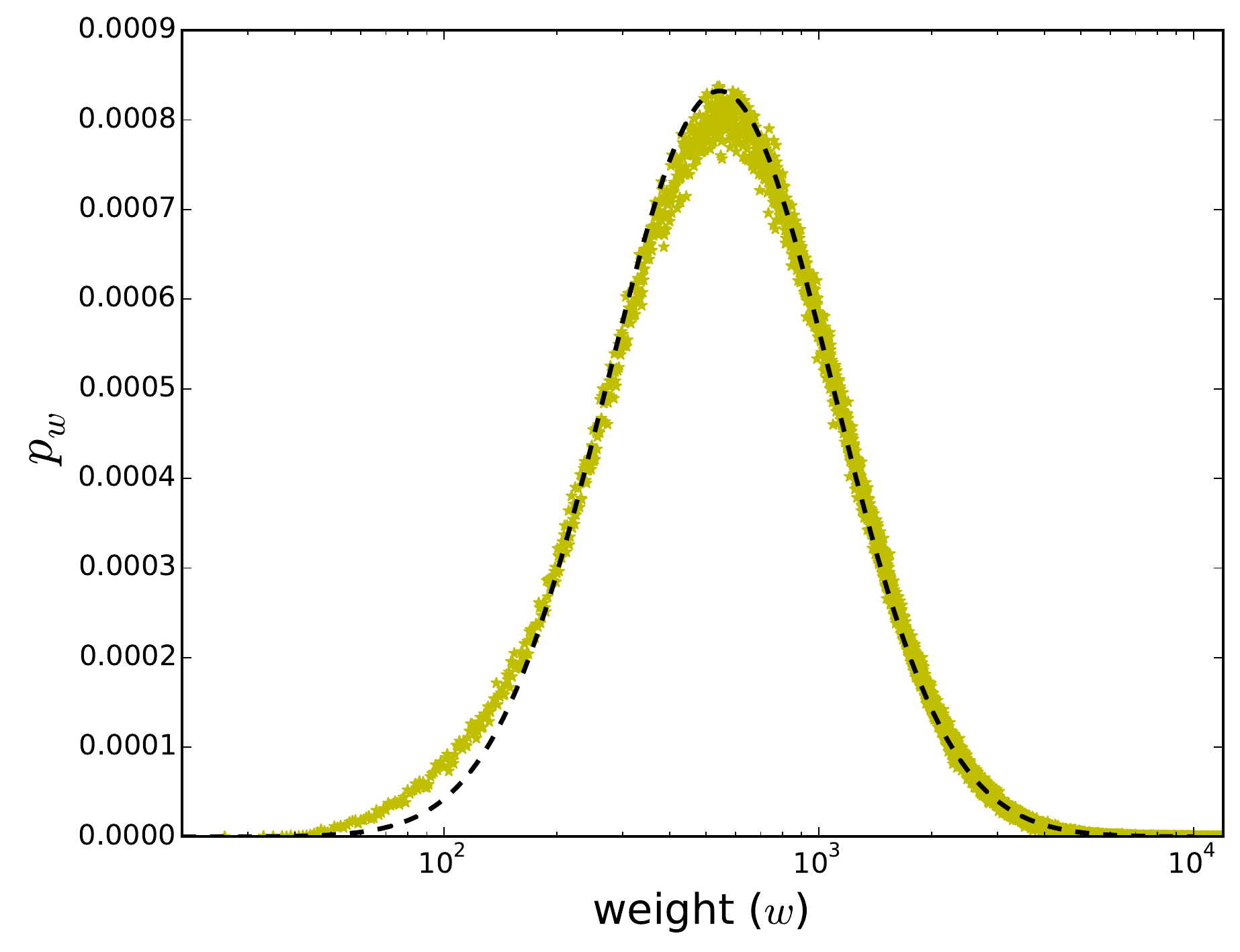}
\caption{On the left panel: The degree distribution and a log-normal fit. The inset shows users with degree $k\geq 100$ and a similar fit. On the right panel: The weighted degree distribution 
is shown 
with a fitted log-normal distribution.} \label{degree_dist}
\end{center}
\end{figure*}

\section{Results}

Although the mobile phone call dataset we study here 
contains almost complete data on over 6 million users, only a fraction of these have a degree $ k \geq 100$. In order to test the hypothesis of the layers of different levels of emotional closeness, we 
analyse users that are in the lower bound of the outer layer, i.e. have called 100 or more alters. This leaves 
$26,680$ users with a mean number of alters of 129.9 and standard deviation of 37.7. The mean number of calls an ego makes is 3553.8 with a standard deviation of 1894.1.
We also analyse users with $50 < k < 100$ ($N = 301,190$). These have an average weighted degree of 1964.2. 
The remaining users with $k < 50$ have an mean weighted degree of 148.8 indicating they use their phone on average less than once every two days.

The degree distribution and the weighted degree distribution for the entire dataset are shown in fig.~\ref{degree_dist} with log-normal distributions fitted. The inset in the left panel shows the degree distribution for users with degree $k\geq 100$, which follows as similar distribution. Both truncated power-law and log-normal models yield high Akaike weights for the degree distribution but with slightly more support for the log-normal behaviour (a truncated power law is fitted to the dataset in \cite{onnela2007structure}). 
Hence we consider that, of the candidate models, the log-normal model has the highest support for the weighted degree distribution.

Log-normal distributions are associated with multiplicative processes, in contrast with Gaussian distributions which are additive. An important consequence of a log-normal distribution is that in a growth process, the growth rate is independent of the size \citep{sutton1997gibrat}. In terms the degree distribution, this means that the rate of growth of an ego's number of friends is independent of their current number of friends. 
Log-normal distributions are found in many empirical datasets 
\citep{eeckhout2004gibrat,mitzenmacher2004brief,Clauset}.

\begin{figure}
\begin{center}
\includegraphics[width=0.99\columnwidth]{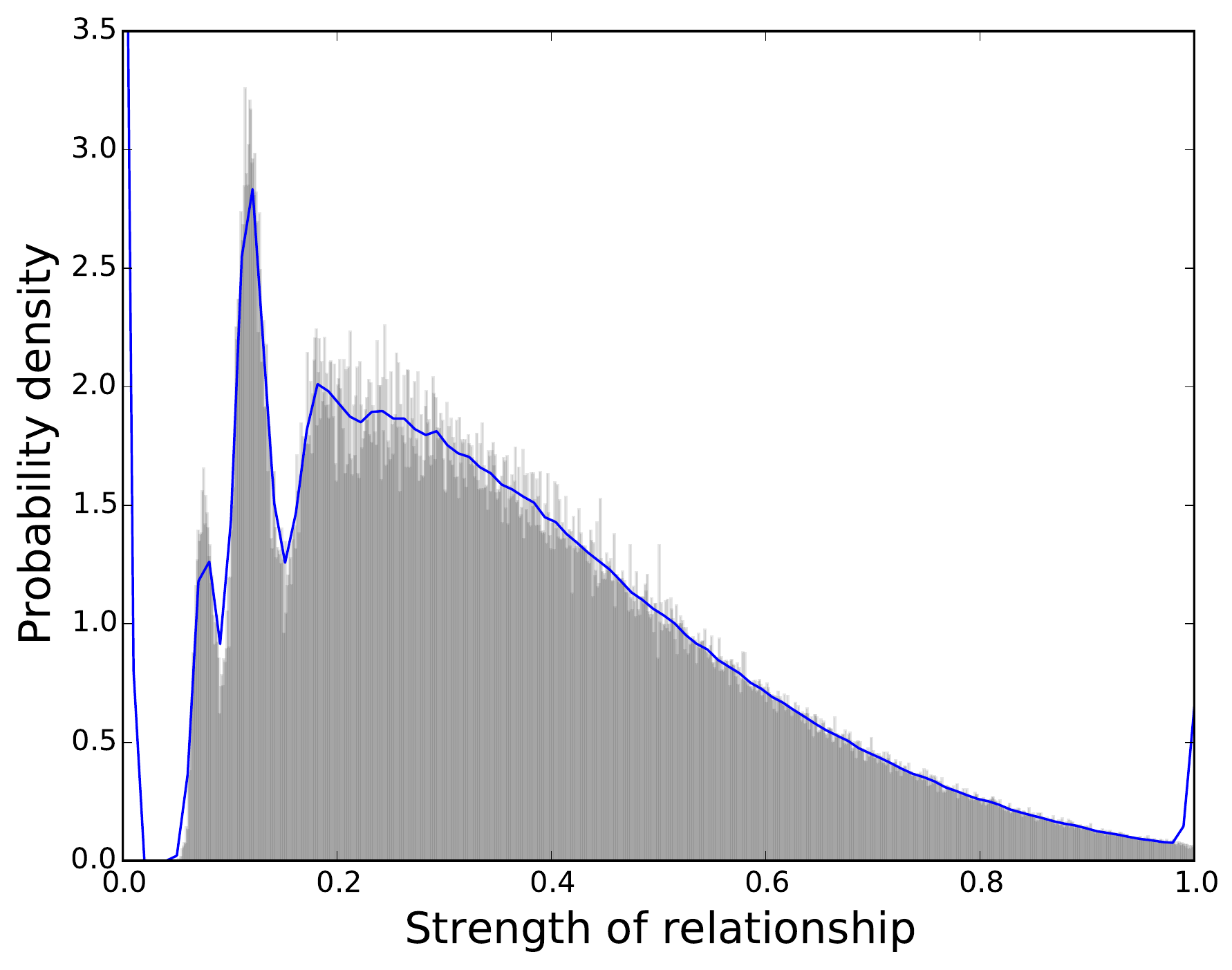}
\caption{The histogram of the normalised weights of each call for all users. The blue line is a Gaussian kernel density estimator to the data.} \label{kde}
\end{center}
\end{figure}

Having established that the data is 
log-normally distributed, before normalising the weight array in eq.~(\ref{eqn:norm}), the log of each weight is taken. Fig.~\ref{kde} shows the histogram for the normalised log of the weights $\hat{w}$ for all users. 
There are peaks at 0 and 1 as every user has at least one alter they call a minimal number of times (usually multiple) and at least one alter they call a maximum number of times. 
Although the data is very noisy, we observe peaks indicating that there are some groupings within the data. The blue line is a Gaussian kernel density estimate of the distribution. The local minima on average split the number of alters into groups of 15 until the normalised strength of a relationship $\hat{w} = 0.28$ after which there are an average of 68.8 alters. This could roughly correspond to the third Dunbar layer of 50. Beyond this region, the data is too noisy to split up further on a group level.

Assuming then 
that there are some kinds of groupings within each user's call data $W$, we next use more traditional clustering algorithms to attempt to identify the layers. Here we do not need to normalise $W$ as we split each user's weight array individually and analyse the overall distributions one by one. 

\begin{figure}
\begin{center}
\includegraphics[width=0.99\columnwidth]{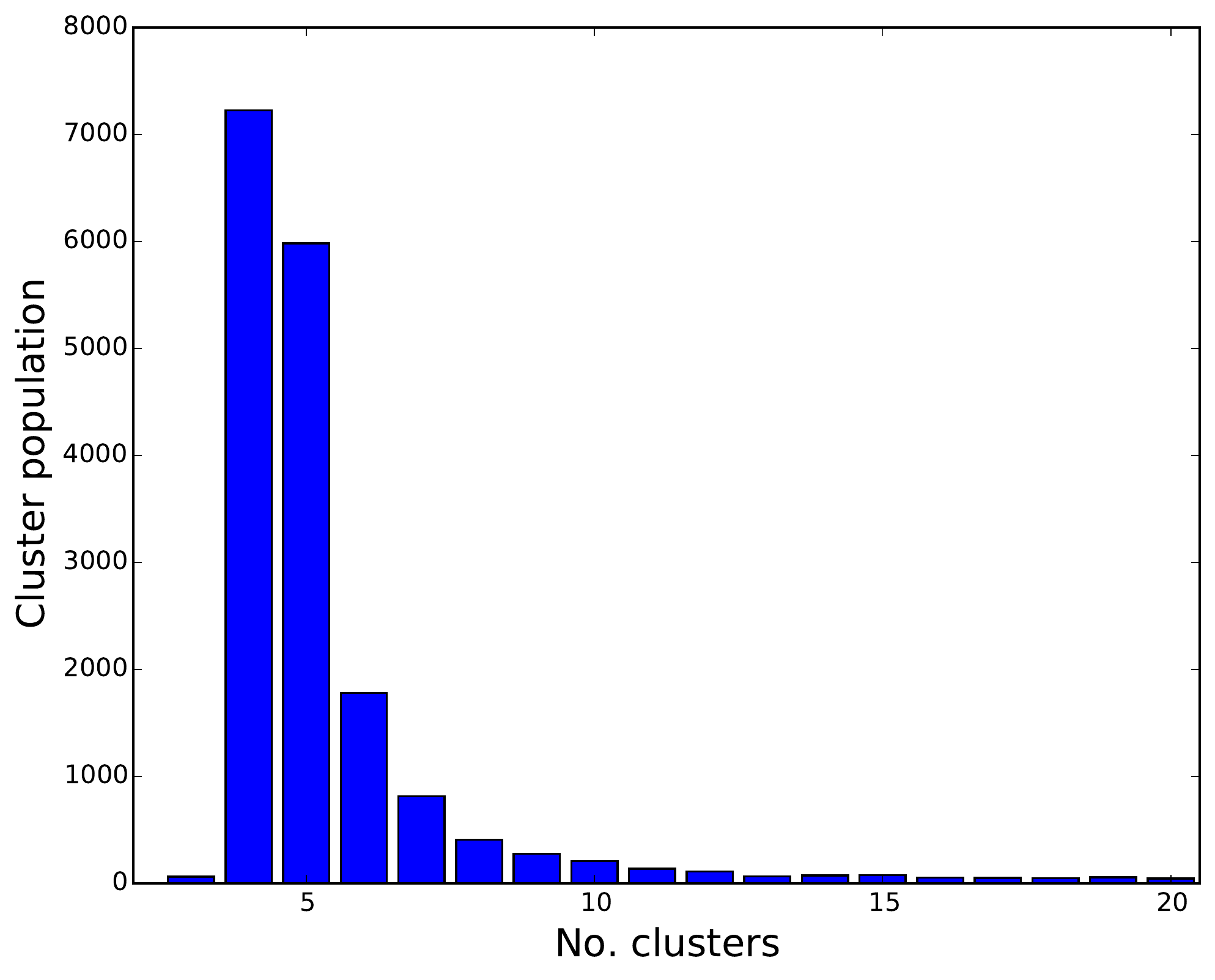}
\caption{Number of users in each cluster using the Jenks algorithm. The majority of users have four layers of friends. The algorithm does not converge well for large numbers ($>7$) of clusters.} \label{jenks_bar}
\end{center}
\end{figure}

The analysis with the Jenks algorithm is found to split almost half of the users (13,209) into 4-5 clusters, but it also finds of the order of a hundred users in every cluster from clusters of 7 and up. 
Fig.~\ref{jenks_bar} shows the number of users in each cluster. 

We identify the most common number of clusters to be 4 for 7,226 ($27.1\%$) users. 
The average number of users and their standard deviation in each cluster are reported in table~\ref{tab:clustering} and the average cumulative layer turns out to hold  $4.1, 11.0, 29.8$ and $128.9$ users. 
These numbers are a little smaller than the conventional numbers for Dunbar layers, but within their natural range of variation.
The next most common number of clusters, as shown in fig.~\ref{jenks_bar}, is 5 clusters for $5,983$ ($22.4\%$) users with 
cumulative layers holding $2.9, 7.4, 17.7, 43.0,$ and 134.3 users. These numbers are quite similar to the Dunbar numbers, but 
with an another layer between the first two.

\begin{figure*}
\begin{center}
\includegraphics[width=0.99\columnwidth]{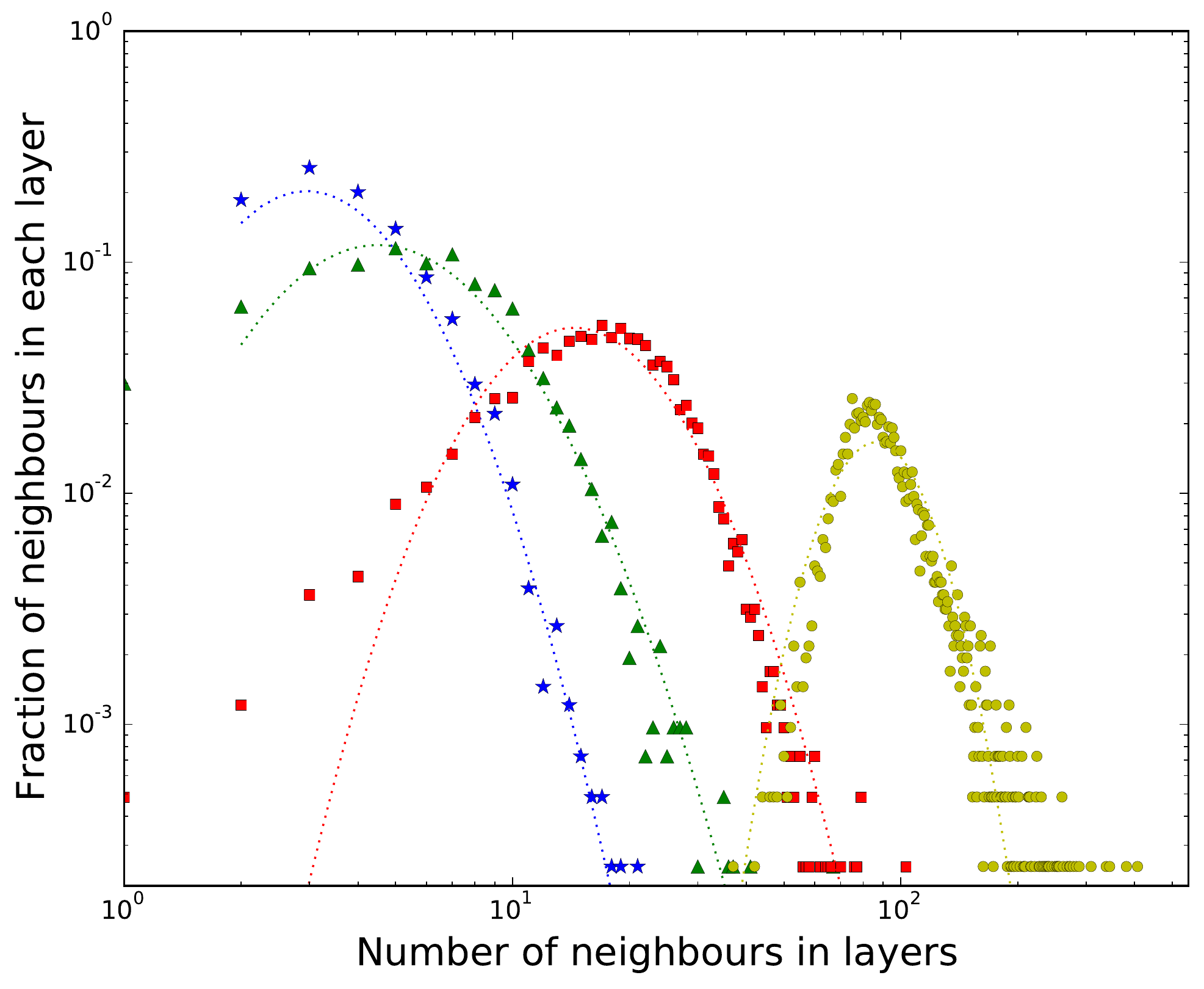}
\includegraphics[width=0.99\columnwidth]{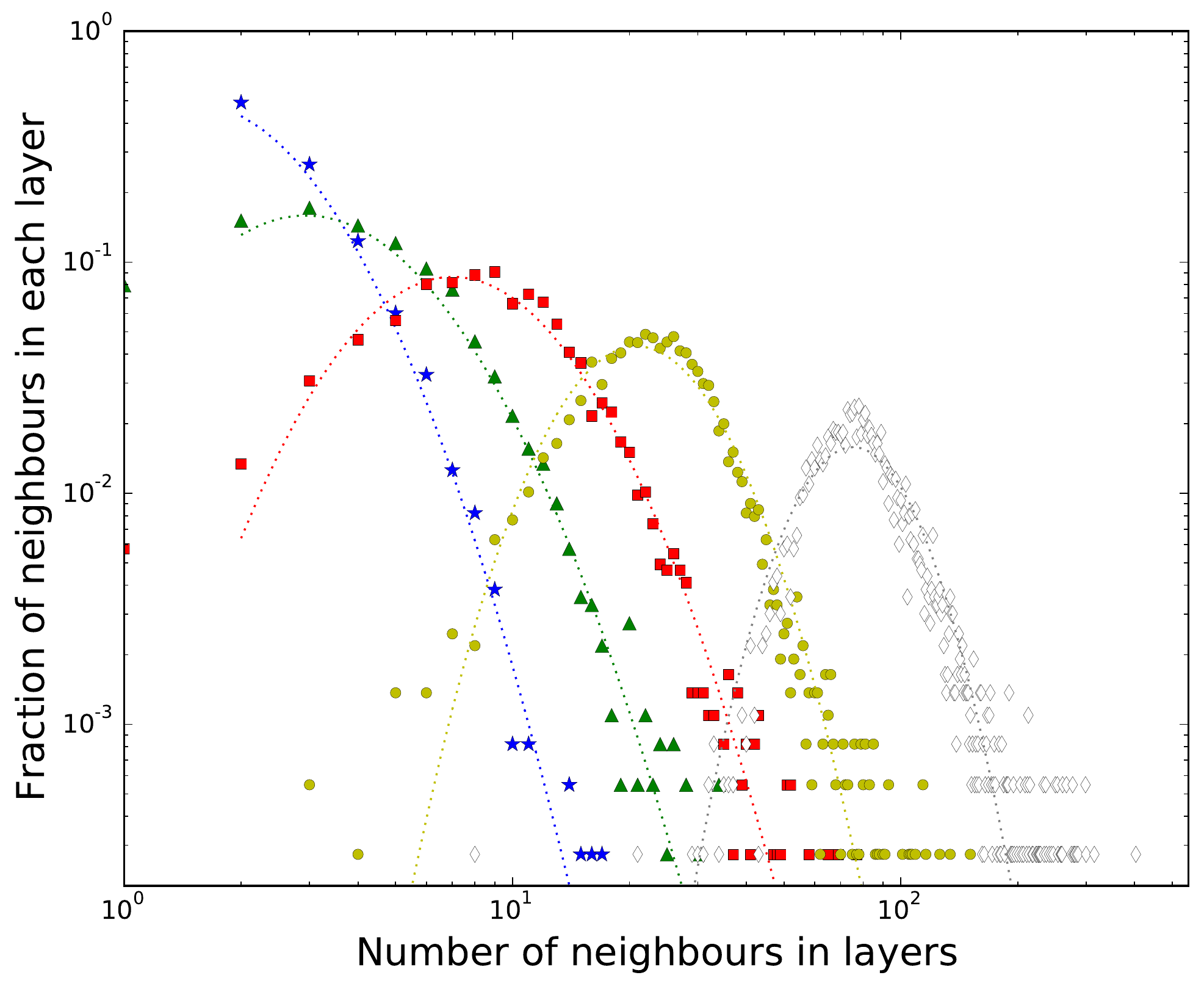}
\caption{The probability distributions for the number of users in each cluster for 4 clusters on the left and 5 clusters on the right using the Jenks algorithm and shown on a log-log scale. Log-normal distributions can be fitted to each as seen by the dotted lines.} \label{jenks_layers}
\end{center}
\end{figure*}

The probability distributions for 4 and 5 clusters are shown in fig.~\ref{jenks_layers}. Each cluster is log-normally distributed. This is perhaps not surprising given that the original distribution is log-normal; there is already a large variation with some users perhaps having 100 friends split into four clusters and others having almost 500 split into four clusters.  
Therefore even if the estimates for the mean size of each layer are accurate, 
there is large variance with numbers of alters going far to the right of these means.

Next we use a Gaussian Mixture Model to split the data. For this algorithm, we however take the log of the data as it is log-normally distributed and not Gaussian. 
In this case 9,289 of the users are split into 2 clusters and 3,334 users into 3 clusters. (The model finds about 1000 thousand users in each cluster above this up to 12 clusters.) The average and standard deviations are reported in table~\ref{tab:clustering}.
For 2 clusters, the cumulative layer-size means are 43.0 and 122.9 users, which again are quite close to the outer two Dunbar layers. For 3 clusters the means are at 18.0, 59.7 and 126.4 users which are close to the outer three layers but miss the inner one.

Lastly we apply the head/tail breaks algorithm. For users with more than 100 alters, it finds 12,951 users in 3 clusters with cumulative layer sizes of 4.3, 22.4 and 122.9 users and 11,616 users in 4 clusters with cumulative layer sizes of 2.2, 8.2, 29.8 and  133.5 users. The full details are reported in table~\ref{tab:clustering}. Again, the layers are found to be log-normally distributed using this method. 

Each of the three algorithms finds small number of layers for the majority of the data. The Jenks and head/tails algorithms find an inner layer ranging from 2.2 to 4.3 alters and almost all algorithms yield a large outer layer of 80-100 alters at the end. The Jenks algorithm and Gaussian Mixture Model both give good evidence for the outer two layers.

If we assume that the majority of 4 clusters is the appropriate value, we can use the Jenks algorithm to force everyone into 4 clusters. The values of the 
cumulative layers are then 3.5, 10.6, 31.1 and 129.9 users, which are slightly smaller numbers than those conventionally considered for the Dunbar layers, but have virtually the same scaling ratio between the layers, i.e. on average, 3.3 here, compared to an average of 3.2 found by \cite{zhou2005discrete} and 3.2 in Facebook traffic by \cite{dunbar2015structure}.

These results are from call frequency. We also have the duration of each call. The duration and call frequency are highly correlated with a Pearson correlation coefficient of $ r= 0.71$. For egos with $k \geq 100$ this is slightly higher at $r=0.76$. Using the Jenks algorithm this again yields a majority of egos with 4 clusters (27\% of egos) with cumulative layers of 3.9, 10.1, 27.2 and 129.3. These are slightly lower than the results for call frequency. 

Finally we apply the Jenks algorithm to egos' call frequencies with $50 \leq k < 100$ ($N =301,190 $). In this case the majority of users are found to be best clustered into three (23.5\%) with cumulative layers of 3.9, 11.9 and 63.91. This matches the first three Dunbar layers well. The next most common is 4 clusters (14.5\%) with cumulative layers of 2.7, 6.7, 17.9 and 64.3. These also contain the first three Dunbar layers well but have an additional lower value close to the 1.5 layer found in \cite{dunbar2015structure}. Once again the duration gives very similar results to the call frequency.

\begin{table*} 
\begin{center}
 \caption{The average number of users in each cluster using the Jenks Natural Breaks algorithm, Gaussian Mixture Model (GMM) and the Head/Tail Breaks (H/T). The number of users in a cluster is given by $N$, the total number of clusters the algorithm finds is denoted by $c$ and $n_i$ gives the average number of alters in cluster $i$. The cumulative number of users in each cluster is given in the second part of the table}
 \label{tab:clustering}
\small
\centering
  \begin{tabular*}{.99\textwidth}{@{\extracolsep{\fill} }l|c c c c c c c}
    & $N$ & $n_1$ & $n_2$ &$n_3$ & $n_4$&  $n_5$ 
     \\ \hline
  Jenks\\
  $c=4$  & 7226 &  4.1  (2.0)  &  6.8 (4.0) &  18.9 (8.7)  &  99.1 (32.7)  \\
   $c = 5$ & 5983 & 2.9 (1.3)  & 4.6 (3.2)  &  10.2 (6.1)  &  25.4 (11.4)   &  91.3 (32.7) \\
 \hline \hline
   GMM \\ 
      $c=2$ & 9289 & 43.0 (15.5) & 79.9 (21.3) \\
      $c=3$ & 3334 & 18.0 (10.1) & 41.8 (15.4) & 66.6 (19.8)  \\
   H/T & \\
   $c=3$ & 12951 & 4.3 (1.9)  & 18.1 (5.5)  & 98.5 (22.9) \\
   $c=4$ & 11616 & 2.2 (0.9) & 5.9 (2.5) & 21.6 (7.7) & 103.7 (32.9) \\
 \hline \hline  \\
  \end{tabular*}
  
  Cumulative
  
   \begin{tabular*}{.99\textwidth}{@{\extracolsep{\fill} }l|c c c c c c c}
     & $N$ & $n_1$ & $n_2$ &$n_3$ & $n_4$&  $n_5$ \\ \hline
   Jenks\\
   $c=4$  & 7226 & $ 4.1 $ & 11.0 & 29.8 & 128.9 \\
    $c = 5$ & 5983 & $ 2.9 $ & 7.4 & 17.7 & 43.0 & 134.3\\
  \hline \hline
    GMM \\ 
       $c=2$  & 9289  & 43.0 & 122.9  \\
    $c=3$ & 3334 & 18.0 & 59.7 & 126.4 \\ \hline \hline
    H/T & \\
    $c=3$ & 12951 & $4.3$ & 22.4 & 120.9 \\
    $c=4$ & 11616 & $2.2$ & 8.2 & 29.8 & 133.5 \\
  \hline \hline   
   \end{tabular*} 
\end{center}
\end{table*}

\section{Conclusions}

In this study, by applying different clustering algorithms to a mobile phone dataset we find strong evidence for a layering structure.
Fig.~\ref{kde}, for example, makes no prior assumptions and shows that there is some
structure within the dataset in spite of all the noise.
However, finding discrete layers is still a considerable challenge.

Although the clustering methods yield slightly different results, as shown 
at the bottom of table~\ref{tab:clustering}, they have important similarities. They all find a small number of clusters and show good support for the outer two layers. While the data is noisy, all methods support two different groupings well. 
 This could, for example, mean introverts and extroverts have a different number of layers of friends. Further work could investigate this possibility.
Another suggestion is that
over a year, friendships are more transient. Alters could move up or down from one layer to the next on a regular basis. This would reflect the temporal nature of emotional closeness, especially among one's non-closest friends.

The Jenks algorithm and Gaussian Mixture Model for 4 layers, they give results close to the Dunbar layers. 
In addition, they have the same scaling pattern ($\sim 3$) as has previously been reported for the structure of offline egocentric social networks and the organisation of natural communities \citep{zhou2005discrete,hamilton2007complex}, Facebook and Twitter traffic \citep{dunbar2015structure} and online gaming environments \citep{fuchs2014fractal}. We still do not have any principled explanation for why these structural layers  should have such a consistent pattern, but they are closely tied in to the psychological aspects of relationships like emotional closeness \citep{sutcliffe2012relationships}. 

The means in the clusters are generally smaller than the predicted means from the Dunbar layers, though they match the ranges found in \cite{hill2003social}. A reason they could be smaller here is due to the fact that a mobile phone call dataset only captures a portion of an ego's social network, even with taking users who call more than 100 alters. With many other modes of communication available, it is unlikely that a user would only resort to phoning their friends. We emphasise however, that the year 2007 is a good time to use cellphone data for this kind of analysis as it is just prior to smart phones (the first \emph{iPhone} was released a few months before this dataset ends) which facilitates many other 
avenues of online communication.
It is also before platforms, such as Skype or Facebook, were at the height of their popularity.
We also show that users who call 50 to 100 people throughout the year match the inner three Dunbar layers well. 

This study has strong implications for the social brain hypothesis as, regardless of the mode of communication, similar structure is observed. Future work on this will involve analysing the turnover time in the layers using temporal data in different communication datasets. We believe that the turnover time will relate to the emotional closeness, for example the inner layer is likely to be more robust in phone data than on Twitter.

A final point of note here is on the structure of the layers. From each algorithm we find that the layers are log-normally distributed for all number of clusters. This, to the best of our knowledge, has not been observed before. 
It is important to point out that this rightward skew and large standard deviation exist. 
The log-normality is due to the entire degree distribution being log-normal and thus already having considerable skew. 
This potentially also shows a difference between extroverts, who tend to have a number of friends far greater than the mean. Regardless of their number of friends however, they still show evidence of layers.

\begin{acknowledgments}
The authors would like to thank Hang-Hyun Jo for helpful discussions. P.MC. and R.D. are supported by a European Research Council (No. 295663) Advanced Investigator grant to R.D. .K.K. acknowledges financial support from the Academy of Finland's COSDYN project (no. 276439).
\end{acknowledgments}

\bibliography{mobile_phone_bib}

\end{document}